\documentclass[10pt]{article}
\usepackage{amsmath}
\usepackage{latexsym}
\usepackage{amssymb}
\usepackage{amsfonts}
\usepackage{amsthm}
\title{A MATHEMATICAL COMMENT \\ ON GRAVITATIONAL WAVES}
\author{J.-F. Pommaret \\ CERMICS, Ecole des Ponts ParisTech,\\ 6/8 Av. Blaise Pascal, 77455 Marne-la-Vall\'ee Cedex 02, France \\
E-mail: jean-francois.pommaret@wanadoo.fr\\
URL: http://cermics.enpc.fr/$\sim$pommaret/home.html }
\date{  }
\begin{document}
\maketitle

\noindent

{\bf ABSTRACT}  \\

In classical General Relativity, the way to exhibit the equations for the gravitational waves is based on two " {\it tricks} " allowing to transform the Einstein equations after linearizing them over the Minkowski metric. With specific notations used in the study of {\it Lie pseudogroups} of transformations of an $n$-dimensional manifold, let $\Omega=({\Omega}_{ij}={\Omega}_{ji})$ be a perturbation of the non-degenerate metric $\omega=({\omega}_{ij}={\omega}_{ji})$ with $det(\omega)\neq 0$ and call ${\omega}^{-1}=({\omega}^{ij}={\omega}^{ji})$ the inverse matrix appearing in the Dalembertian operator $\Box = {\omega}^{ij}d_{ij}$. The first idea is to introduce the linear transformation ${\bar{\Omega}}_{ij}={\Omega}_{ij}-\frac{1}{2}{\omega}_{ij}tr(\Omega)$ where $tr(\Omega)={\omega}^{ij}{\Omega}_{ij}$ is the {\it trace} of $\Omega$, which is invertible when $n\geq 3$. The second important idea is to notice that the composite second order linearized {\it Einstein} operator $\bar{\Omega} \rightarrow \Omega \rightarrow E=(E_{ij}=R_{ij} - \frac{1}{2}{\omega}_{ij}tr(R))$ where $\Omega \rightarrow R=(R_{ij}=R_{ji})$ is the linearized {\it Ricci} operator with trace $tr(R)={\omega}^{ij}R_{ij}$ is reduced to $\Box {\bar{\Omega}}_{ij}$ when ${\omega}^{rs}d_{ri}{\bar{\Omega}}_{sj}=0$. The purpose of this short but striking paper is to revisit these two results in the light of the {\it differential duality} existing in {\it Algebraic Analysis}, namely a mixture of differential geometry and homological agebra, providing therefore a totally different interpretation. In particular, we prove that the above operator $\bar{\Omega} \rightarrow E$ is nothing else than the formal adjoint of the {\it Ricci} operator $\Omega \rightarrow R$ and that the map $\Omega \rightarrow \bar{\Omega}$ is just the formal adjoint (transposed) of the defining tensor map $R \rightarrow E$. Accordingly, the {\it Cauchy} operator (stress equations) can be directly parametrized by the formal adjoint of the {\it Ricci} operator and the {\it Einstein } operator is no longer needed. \\

\vspace{4cm}

\noindent
{\bf KEY WORDS}  \\
General Relativity, Killing equations, Ricci tensor, Einstein tensor, Conformal Killing equations, Weyl tensor, Lie group, Lie pseudogroup, Algebraic Analysis, Homological algebra, Differential duality, Adjoint operator.  \\

\newpage

{\bf 1) INTRODUCTION}  \\

In order to make the paper rather self-contained, we recall a few notations and definitions on linear systems of partial differential (PD) equations [8-12,22,23,28]. If $E$ is a vector bundle over the base manifold $X$ with projection $\pi$ and local coordinates $(x,y)=(x^i,y^k)$ projecting onto $x=(x^i)$ for $i=1,...,n$ and $k=1,...,m$, identifying a map with its graph, a (local) section $f:U\subset X \rightarrow E$ is such that $\pi\circ f =id$ on $U$ and we write $y^k=f^k(x)$ or simply $y=f(x)$. For any change of local coordinates $(x,y)\rightarrow (\bar{x}=\varphi(x),\bar{y}=A(x)y)$ on $E$, the change of section is $y=f(x)\rightarrow \bar{y}=\bar{f}(\bar{x})$ such that ${\bar{f}}^l(\varphi(x)\equiv A^l_k(x)f^k(x)$. The new vector bundle $E^*$ obtained by changing the {\it transition matrix} $A$ to its inverse $A^{-1}$ is called the {\it dual vector bundle} of $E$. We may introduce the tangent bundle $T$, the cotangent bundle $T^*$, the vector bundle $S_qT^*$ of $q$-symmetric covariant tensors and the vector bundle ${\wedge}^rT^*$ of $r$-skewsymmetric covariant tensors or $r$-forms. Differentiating with respect to $x^i$ and using new coordinates $y^k_i$ in place of ${\partial}_if^k(x)$, we obtain ${\bar{y}}^l_r{\partial}_i{\varphi}^r(x)=A^l_k(x)y^k_i+{\partial}_iA^l_k(x)y^k$. Introducing a multi-index $\mu=({\mu}_1,...,{\mu}_n)$ with length $\mid \mu \mid={\mu}_1+...+{\mu}_n$ and prolonging the procedure up to order $q$, we may construct in this way a vector bundle $J_q(E)$ over $X$, called the {\it jet bundle of order} $q$ with local coordinates $(x,y_q)=(x^i,y^k_{\mu})$ with $0\leq \mid\mu\mid \leq q$ and $y^k_0=y^k$. For a later use, we shall set $\mu+1_i=({\mu}_1,...,{\mu}_{i-1},{\mu}_i+1,{\mu}_{i+1},...,{\mu}_n)$ and define the operator $j_q:E \rightarrow J_q(E):f \rightarrow j_q(f)$ on sections by the local formula $j_q(f):(x)\rightarrow({\partial}_{\mu}f^k(x)\mid 0\leq \mid\mu\mid \leq q,k=1,...,m)$. Finally, as the background will always be clear enough, we shall use the same notation for a vector bundle and its set of sections.   \\

\noindent
{\bf DEFINITION 1.1}:  A {\it system} of PD equations of order $q$ on $E$ is a vector subbundle $R_q\subset J_q(E)$ locally defined by a constant rank system of linear equations for the jets of order $q$ of the form $ a^{\tau\mu}_k(x)y^k_{\mu}=0$. Its {\it first prolongation} $R_{q+1}\subset J_{q+1}(E)$ will be defined by the equations $ a^{\tau\mu}_k(x)y^k_{\mu}=0, a^{\tau\mu}_k(x)y^k_{\mu+1_i}+{\partial}_ia^{\tau\mu}_k(x)y^k_{\mu}=0$ which may not provide a system of constant rank. A system $R_q$ is said to be {\it formally integrable} if the $R_{q+r}$ are vector bundles $\forall r\geq 0$ ({\it regularity condition}) and no new equation of order $q+r$ can be obtained by prolonging the given PD equations more than $r$ times, $\forall r\geq 0$. The {\it symbols} $g_{q+r}=R_{q+r}\cap S_{q+r}T^*\otimes E$ only depend on 
$g_q$ [8-12,28].\\

\noindent
{\bf DEFINITION 1.2}: Considering the short exact sequence $0 \rightarrow R_q \rightarrow J_q(E) \stackrel{\Phi}{\longrightarrow} F_0 \rightarrow 0 $ where $\Phi : j_q(E) \rightarrow J_q(E)/R_q$ is the canonical projection, we may thus introduce the linear operator ${\cal{D}}=\Phi \circ j_q:E \rightarrow F_0$. However, as $F_0$ is only defined up to an isomorphism, things may not be so simple when $q=1$ and there is no zero order PD equations. We have the commutative and exact diagram:  \\
\[ \begin{array}{rcccccl} 
  &  0  &  &  0  &  &  0  &  \\
  &  \downarrow  &  &  \downarrow  &  &  \downarrow &   \\
 0  \rightarrow & g_1  &  \rightarrow & T^*\otimes E &  \stackrel{\sigma(\Phi)}{\longrightarrow} & F_0  &  \rightarrow 0  \\ 
  &  \downarrow  &  &  \downarrow  &  & \parallel  &   \\
0  \rightarrow &  R_1  &  \rightarrow  & J_1(E)  &  \stackrel{\Phi}{\longrightarrow} & F_0 &  \rightarrow 0  \\
  &  \downarrow  &  &  \downarrow  &  &  \downarrow &   \\
0  \rightarrow & E  & =  &  E  &  \rightarrow 0  &  \\
 &  \downarrow  &  &  \downarrow  &  &  &   \\      
  &  0  &  &  0  &  &  &
 \end{array}  \]
where $\sigma(\Phi)$ is the induced symbol epimorphism.  \\

\noindent
{\bf EXAMPLE 1.3}: The infinitesimal isometries of the non-degenerate metric $\omega\in S_2T^*$ with $det(\omega)\neq 0$ are defined by the kernel $\Theta$ of the linear first order {\it  Killing operator} $T \rightarrow S_2T^*:\xi \rightarrow {\cal{D}}\xi={\cal{L}}(\xi)\omega=\Omega  $, which involves the Lie derivative ${\cal{L}}$ and provides twice the so-called infinitesimal deformation tensor of continuum mechanics when $\omega$ is the Euclidean metric. We may consider the linear first order system of {\it general infinitesimal Lie equations} in {\it Medolaghi form}, also called {\it system of Killing equations} [8,11,29]:  \\
\[{\Omega}_{ij}\equiv ({\cal{L}}(\xi){\omega})_{ij} \equiv {\omega}_{rj}(x){\partial}_i{\xi}^r+{\omega}_{ir}(x){\partial}_j{\xi}^r+{\xi}^r{\partial}_r{\omega}_{ij}(x)=0 \]
which is in fact a family of systems only depending on the {\it geometric object} $\omega$ and its derivatives. Introducing the Christoffel symbols $\gamma$, we may differentiate once and add the operator ${\cal{L}}(\xi)\gamma=\Gamma \in S_2T^*\otimes T$ with the well known Levi-Civita isomorphism $j_1(\omega)=(\omega,{\partial}_x\omega)\simeq (\omega,\gamma)$ in order to obtain the linear second order system of {\it general infinitesimal Lie equations} in {\it Medolaghi form}:  \\
\[ {\Gamma}^k_{ij}\equiv ({\cal{L}}(\xi)\gamma)^k_{ij}\equiv {\partial}_{ij}{\xi}^k+{\gamma}^k_{rj}(x){\partial}_i{\xi}^r+{\gamma}^k_{ir}(x){\partial}_j{\xi}^r-{\gamma}^r_{ij}(x){\partial}_r{\xi}^k+{\xi}^r{\partial}_r{\gamma}^k_{ij}(x)=0   \]
This system is formally integrable if and only if $\omega$ has a {\it constant Riemannian curvaure} [2,8-11]. In the diagram, $E=T, F_0=S_2T^*$ and $\sigma(\Phi):T^*\otimes T \rightarrow S_2T^*:{\xi}^k_i \rightarrow {\omega}_{rj}(x){\xi}^r_i+{\omega}_{ir}(x){\xi}^r_j$.  \\
Similarly, introducing the Jacobian determinant $\Delta (x)=det({\partial}_if^k(x))$ and the {\it metric density} ${\hat{\omega}}_{ij}={\mid det(\omega)\mid }^{-\frac{1}{n}}{\omega}_{ij} \Rightarrow det(\hat{\omega})=\pm 1$ as a new {\it geometric object}, rather than by eliminating a {\it conformal factor} as usual, the infinitesimal {\it conformal isometries} are defined by the kernel $\hat{\Theta}$ of the {\it conformal Killing operator} $\xi \rightarrow \hat{\cal{D}}\xi={\cal{L}}(\xi)\hat{\omega}=\hat{\Omega }$. We may consider the first order system of {\it general infinitesimal Lie equations} in {\it Medolaghi form}, also called {\it system of conformal Killing equations} [16,17]:  \\
\[ { \hat{\Omega}}_{ij}\equiv {\hat{\omega}}_{rj}(x){\partial}_i{\xi}^r +{\hat{\omega}}_{ir}(x){\partial}_j{\xi}^r 
- \frac{2}{n}{\hat{\omega}}_{ij}(x){\partial}_r{\xi}^r + {\xi}^r{\partial}_r{\hat{\omega}}_{ij}(x)=0    \]
With first prolongation obtained by eliminating the arbitrary $1$-form $(A_i(x)dx^i)\in T^*$ in:  \\
\[  ({\cal{L}}(\xi)\gamma)^k_{ij}= {\delta}^k_iA_j+{\delta}^k_jA_i-{\omega}_{ij}{\omega}^{kr}A_r   \]
We may introduce the {\it trace} $tr(\Omega)={\omega}^{ij}{\Omega}_{ij}$ with standard notations and obtain therefore $tr(\hat{\Omega})=0$ because ${\hat{\Omega}}_{ij}={\mid det(\omega)\mid }^{-\frac{1}{n}} ({\Omega}_{ij}- \frac{1}{n}{\omega}_{ij}tr(\Omega))$ by linearization. This system is formally integrable if and only if the corresponding {\it Weyl tensor} vanishes [8,9,10]. In the diagram $E=T, {\hat{F}}_0=\{ \hat{\Omega}\in S_2T^*\mid tr(\hat{\Omega})=0 \}$ and $ \sigma(\hat{\Phi}):T^*\otimes T\rightarrow {\hat{F}}_0: {\xi}^k_i \rightarrow {\omega}_{rj}(x){\xi}^r_i+{\omega}_{ir}(x){\xi}^r_j - \frac{2}{n}{\omega}_{ij}{\xi}^r_r$.  \\
The inclusions $R_1\subset{\hat{R}}_1\Rightarrow g_1 \subset {\hat{g}}_1$ induces an epimorphism $F_0 \rightarrow {\hat{F}}_0$ described by 
$ {\Omega}_{ij}\rightarrow {\hat{\Omega}}_{ij}={\Omega}_{ij} - \frac{1}{n}{\omega}_{ij}tr(\Omega)$. Contrary to the Abstract, {\it this is the only combination having a purely mathematical meaning related to group theory but never invertible}. It is only in the next Section that we shall understand the origin of this confusing fact.  \\

Prolonging twice the the first diagram of this paper while using only the symbol top rows, we get the following commutative diagram where all the sequences are exact but the left column:  \\
\[  \begin{array}{rcccccccl}
   & 0  &  &  0  &  &  0  &  &  &  \\
   & \downarrow & & \downarrow & & \downarrow & & &  \\
   0 \rightarrow & g_3 & \longrightarrow & S_3T^*\otimes E & \longrightarrow & S_2T^*\otimes F_0 & \longrightarrow &F_1  & \rightarrow 0  \\
  & \hspace{3mm}  \downarrow \delta & & \hspace{3mm}   \downarrow \delta  & & \hspace{3mm}\downarrow \delta & & &  \\   
0 \rightarrow &  T^*\otimes g_2  & \longrightarrow & T^*\otimes S_2T^*\otimes E & \longrightarrow  &  T^*\otimes T^*  \otimes F_0  & \longrightarrow & 0  &   \\
  & \hspace{3mm}  \downarrow \delta & & \hspace{3mm}   \downarrow \delta  & & \hspace{3mm}\downarrow \delta   & & &  \\   
0\rightarrow & {\wedge}^2T^*\otimes g_1 &  \longrightarrow & {\wedge}^2T^*\otimes T^*\otimes E  & \longrightarrow & 
{\wedge}^2T^*\otimes F_0 & \longrightarrow &  0  & \\
  &  \hspace{3mm}  \downarrow \delta & & \hspace{3mm}   \downarrow \delta  & & \downarrow & & &  \\  
  0 \rightarrow & {\wedge}^3T^*\otimes E  &  =  &  {\wedge}^3T^*\otimes E &  \longrightarrow &  0  & &  &  \\
   &  \downarrow &   &  \downarrow  & & & & &  \\
    &  0 &   & 0 & & & & &  \\  
\end{array}  \]
In the situations considered, we have $E=T$, $g_1\subset T^*\otimes T \Rightarrow g_2=0 \Rightarrow  g_3=0  \Rightarrow $ and $ {\hat{g}}_1\subset T^*\otimes T \Rightarrow {\hat{g}}_2\simeq T^* \Rightarrow {\hat{g}}_3=0$, a result leading to $F_1\simeq H^2_1(g_1)$ and ${\hat{F}}_1\simeq H^2_1({\hat{g}}_1)$.  \\  

We have explained in books [8-13] or papers [15-23] how to construct a {\it differential sequence}:   \\
\[  0  \rightarrow \Theta \rightarrow T  \stackrel{\cal{D}}{\longrightarrow} F_0 \stackrel{{\cal{D}}_1}{\longrightarrow} F_1 \stackrel{{\cal{D}}_2}{\longrightarrow} F_2 \rightarrow ...   \] 
where each operator generates the CC of the previous one, ${\cal{D}}$ is first order, ${\cal{D}}_1$ is second order and is the linearization $(R^k_{l,ij})$ of the Riemann tensor over a given flat metric like the Minkowski metric while ${\cal{D}}_2$ is again first order and is the linearization of the Bianchi identities with:  \\
\[  \begin{array}{rcl}
 dim(F_1) & = &  dim(S_2T^*\otimes F_0)- dim(S_3T^*\otimes T)  \\
                & = &  dim({\wedge}^2T^*\otimes g_1) - dim({\wedge}^3T^*\otimes T)  \\
                & = & n^2(n^2-1)/12 
                \end{array}   \]
while ${\cal{D}}_2$ is again first order and is the linearization of the Bianchi identities with:  \\
\[  \begin{array}{rcl}
dim(F_2)  & = & dim( S_4T^*\otimes T) - dim(S_3T^*\otimes F_0) + (dim(T^*\otimes F_1)  \\
                & = & dim ({\wedge}^3T^*\otimes g_1) - dim({\wedge}^4T^*\otimes T)  \\
               &  =  &  n^2(n^2-1)(n-2)/24 
        \end{array}   \]
         
{\it The conformal situation is drastically different} but not acknowledged today, because ${\hat{g}}_3=0, \forall n\geq 3$ and we have to study {\it separately} the cases $n=3,n=4, n\geq 5$ even though $\hat{\cal{D}}$ is still first order, because ${\hat{\cal{D}}}_1$ is third order when $n=3$ but still second order and is the linearization ${\Sigma}^k_{l,ij}$ of the Weyl tensor when $n\geq 4$ with:  \\
\[   \begin{array}{rcl}
  dim({\hat{F}}_1) & = & dim(S_2T^*\otimes {\hat{F}}_0) - dim(S_3T^*\otimes T)  \\
                            & = & dim({\wedge}^2T^*\otimes {\hat{g}}_1) - dim({\wedge}^3T^*\otimes T) -dim({\wedge}^2T^*\otimes {\hat{g}}_2)\\
                            & = & dim(F_1) - dim(S_2T^*)  \\
                            & =& n(n+1)(n+2)(n-3)/12
  \end{array}    \]
while ${\hat{\cal{D}}}_2$ is first order when $n=3$, second order when $n=4$ but again first order when $n\geq 5$ [20-23]. For $n\geq 4$, we have the commutative and exact diagram:  \\ 
\[ \begin{array}{rcccccl} 
  &   &  &  0  &  &  0  &  \\
  &  &  &  \downarrow  &  &  \downarrow &   \\
  & 0 &  \rightarrow & S_2T^* &  = &S_2T^* &  \rightarrow 0  \\ 
  &  \downarrow  &  &  \downarrow  &  & \,\,\, \downarrow \uparrow &   \\
0  \rightarrow &S_3T^*\otimes T &  \rightarrow  & S_2T^*\otimes F_0 & {\rightarrow} & F_1 &  \rightarrow 0  \\
  &  \parallel &  &  \downarrow  &  & \,\,\, \downarrow \uparrow &   \\
0  \rightarrow & S_3T^*\otimes T & \rightarrow  &  S_2T^*\otimes {\hat{F}}_0  &  \rightarrow & {\hat{F}}_1 &\rightarrow 0  \\
 &  \downarrow  &  &  \downarrow  &  & \downarrow &   \\      
  &  0  &  &  0  &  & 0 &
 \end{array}  \]
providing an epimorphism $F_1 \rightarrow {\hat{F}}_1$ with kernel $S_2T^*$, induced by the epimorphism $F_0 \rightarrow {\hat{F}}_0$ and the relation $dim(F_1) - dim({\hat{F}}_1) =n(n+1)/2$. Using again capital letters for the linearized objects, the central and right columns split with the usual contraction map $F_1\rightarrow S_2T^*:R^k_{l,ij} \rightarrow R^r_{i,rj}=R_{ij}=R_{ji}$ and the tensorial lift ${\hat{F}}_1 \rightarrow F_1: {\Sigma}^k_{l,ij}\rightarrow R^k_{l,ij}$ because ${\Sigma}^r_{l,rj}=0$. However, describing such an elementary diagram while chasing in local coordinates needs a lot of work because no classical technique can be used. With more details, the {\it trace} map $\Omega \rightarrow tr(\Omega)$:  \\
\[   A  \rightarrow  {\Omega}_{ij}=A{\omega}_{ij} \rightarrow \frac{1}{n}tr(\Omega)=A   \]
allows to split the central column as it can be extended by setting:   \\
\[  A_{ij} \rightarrow  {\Omega}_{rs,ij}=A_{ij}{\omega}_{rs} \rightarrow \frac{1}{n}{\omega}^{rs}{\Omega}_{rs,ij}=A_{ij} \]
As it is known that the right column splits [15,20-23], the top isomorphism of the diagram may be described by the jet formulas:  \\
\[     nR_{ij}=(n-2)A_{ij}+{\omega}_{ij}{\omega}^{rs}A_{rs}  \Rightarrow tr(R)=2(n-1){\omega}^{rs}A_{rs}  \]
and the corresponding map which is thus injective is also surjective and we find back {\it exactly} the splitting formulas for the fundamental diagram II of [15,22,23] with ${\tau}_{ij}$ in place of $A_{ij}$ but {\it the identification is not evident because symmetric tensors are replaced by skewsymmetric tensors} like in the preceding formulas allowing to compute dimensions. However, using the diagram of Definition $1.2$, we obtain $ker(F_0 \rightarrow {\hat{F}}_0)\simeq {\hat{g}}_1/g_1\simeq {\wedge}^0T^*$ ($1$ {\it dilatation}). Using the short exact sequence $0 \rightarrow g_2 \rightarrow S_2T^*\otimes T \rightarrow T^*\otimes F_0 \rightarrow 0 $ and a similar sequence for the conformal group, we get $ker(T^*\otimes F_0 \rightarrow T^*\otimes {\hat{F}}_0)\simeq {\hat{g}}_2/g_2\simeq T^*\otimes {\hat{g}}_2\simeq T^*\otimes T^*$ 
($n$ {\it elations}). We may collect these results in the split short exact sequence:  \\
\[  0 \rightarrow S_2T^* \stackrel{\delta}{\longrightarrow} T^*\otimes T^* \stackrel{\delta}{\longrightarrow}  {\wedge}^2T^*  \rightarrow 0  \]
with $(A_{ij}=A_{ji})\rightarrow (A_{i,j}=A_{j,i}) \rightarrow A_{i,j} - A_{j,i}=F_{ij}) $, a result showing that we have the direct sum decomposition:   \\
\[   T^*\otimes {\hat{g}}_2\simeq T^*\otimes T^* \simeq S_2T^*\oplus {\wedge}^2T^*\simeq (A_{ij})\oplus (F_{ij})\simeq (R_{ij}) \oplus (F_{ij})   \]
where $(R_{ij})$ is the {\it Ricci tensor} and $(F_{ij})$ is the {\it electromagnetic field} [10,12,21-23].

Another equivalent approach may be obtained through the following diagram where the rows are exact but only the right column is exact:  \\ 
\[  \begin{array}{rcccccl}
 & & & & & 0 &   \\
 & & & & & \downarrow  &   \\
 & & & 0 & & S_2T^* &  \\
 & & & \downarrow &  & \hspace{3mm} \downarrow \delta &   \\
  & 0  &  \rightarrow  &  T^*\otimes {\hat{g}}_2 & = & T^*\otimes T^*  &  \rightarrow  0  \\
  &  \downarrow &  & \hspace{3mm} \downarrow \delta  &  &  \hspace{3mm} \downarrow \delta &     \\
  0 \rightarrow &  {\wedge}^2T^*\otimes g_1  & \rightarrow  &{\wedge}^2T^*\otimes {\hat{g}}_1  & \rightarrow &  {\wedge}^2T^* &  \rightarrow 0   \\   
    & \hspace{3mm} \downarrow \delta  &  &  \hspace{3mm} \downarrow \delta &   &  \downarrow  &    \\

   0  \rightarrow  & {\wedge}^3T^*\otimes T  &   =  &  {\wedge}^3T^*\otimes T  &  \rightarrow  &  0 &  \\
   &  \downarrow  &  &  \downarrow & &  &     \\
   &  0  &  &  0  & & & 
   \end{array}  \]
 A (difficult) chase left to the reader as an exercise provides the split short exact sequence:  \\
 \[   0 \rightarrow  S_2T^* \rightarrow F_1  \rightarrow {\hat{F}}_1 \rightarrow 0   \Rightarrow F_1\simeq S_2T^* \oplus {\hat{F}}_1  \]  
   
 Now, going one step further on in the differential sequence, if the conformal analogue of the Bianchi identities were first order, we should obtain the long exact sequence after one prolongation:  \\
 \[  0 \rightarrow S_4T^*\otimes T  \rightarrow S_3T^*\otimes {\hat{F}}_0 \rightarrow T^*\otimes  {\hat{F}}_1 \rightarrow {\hat{F}}_2 \rightarrow 0  \] 
Applying the Spencer $\delta$-map to each term as we did before, we should obtain a left column that may {\it not} be exact:  \\
\[  0  \rightarrow {\wedge}^2T^*\otimes {\hat{g}}_2 \stackrel{\delta}{ \longrightarrow} {\wedge}^3T^*\otimes {\hat{g}}_1 \stackrel{\delta}{\longrightarrow} {\wedge}^4T^*\otimes T \rightarrow 0  \]
A first (simple) chase proves that the left $\delta$ is injective, a second (much more delicate) proves that the right $\delta$ is surjective while a third (snake type) chase proves that ${\hat{F}}_2$ is isomorphic to the central $\delta$-cohomology $H^3_1({\hat{g}}_1)$ of this sequence. We get:  \\
\[  \begin{array}{rcl}
dim({\hat{F}}_2) & = & dim(S_4T^*\otimes T) - dim(S_3T^*\otimes {\hat{F}}_0) + dim(T^*\otimes {\hat{F}}_1)  \\
                           & = & (dim( {\wedge}^3T^*\otimes {\hat{g}}_1) - dim({\wedge} 4T^*\otimes T)) - dim({\wedge}^2T^*\otimes {\hat{g}}_2)\\
                           & = & n(n^2-1)(n+2)(n-4)/24
  \end{array}    \]
and a contradiction for $n=4$ {\it only}, whenever $n\geq 4$. It is important to notice that {\it indices have never been used}.  \\

\newpage

Finally, keeping $R$ for the linearized Ricci tensor, we recall a few formulas that can be found in most textbooks [3,5,16,22]. We have thus successively ({\it care} to the factor $2$):  \\
\[   2 {\Gamma}^k_{ij}= {\omega}^{kr}(d_i{\Omega}_{rj}+d_j{\Omega}_{ir} - d_r{\Omega}_{ij})=2{\Gamma}^k_{ji}   \]
\[  2 R_{ij}= {\omega}^{rs}(d_{ij}{\Omega}_{rs}+d_{rs}{\Omega}_{ij}-d_{ri}{\Omega}_{sj} - d_{sj}{\Omega}_{ri})= 2R_{ji}  \]
\[ tr(R)= {\omega}^{ij}R_{ij}={\omega}^{ij}d_{ij}tr(\Omega)-{\omega}^{ru}{\omega}^{sv}d_{rs}{\Omega}_{uv}  \]
\[  E_{ij}=R_{ij}- \frac{1}{2}{\omega}_{ij}tr(R)=E_{ji} \Rightarrow tr(E)={\omega}^{ij}E_{ij}= - \frac{(n-2)}{2}tr(R)    \]
\[  2 E_{ij}\equiv {\omega}^{rs}(d_{ij}{\Omega}_{rs}+d_{rs}{\Omega}_{ij}-d_{ri}{\Omega}_{sj}-d_{sj}{\Omega}_{ri})-{\omega}_{ij}({\omega}^{rs}{\omega}^{uv}d_{rs}{\Omega}_{uv}-{\omega}^{ru}{\omega}^{sv}d_{rs}{\Omega}_{uv})=0  \]
and recall the classical computations described in the Abstract:  \\
\[  {\bar{\Omega}}_{ij}={\Omega}_{ij} - \frac{1}{2}{\omega}_{ij}tr(\Omega)  \Leftrightarrow 
{\Omega}_{ij}={\bar{\Omega}}_{ij}- \frac{1}{(n-2)}{\omega}_{ij}tr(\bar{\Omega})   \]
Substituting, we obtain:  \\
\[ \begin{array}{rcl}
 2E_{ij}& =  & \Box {\bar{\Omega}}_{ij} - {\omega}^{rs}d_{ri}{\bar{\Omega}}_{sj}-{\omega}^{rs}d_{sj}{\bar{\Omega}}_{ri}+{\omega}_{ij}{Ê\omega}^{ru}{\omega}^{sv}d_{rs}{\bar{\Omega}}_{uv}  \\
            & = & \Box \bar{\Omega}_{ij}- d_{ri}{\bar{\Omega}}^r_j -d_{rj}{\bar{\Omega}}^r_i + {\omega}_{ij}d_{rs}{\bar{\Omega}}^{rs}
            \end{array}   \]
We notice at once that, apart from the first Dalembertian term, the other terms factor through $d_r{\bar{\Omega}}^r_i$ a result leading to add the {\it differential constraints} $d_r{\bar{\Omega}}^r_i=0$ in a coherent way with the identities:  \\
\[  2{\omega}^{ti}d_tE_{ij}=({\omega}^{ti}{\omega}^{rs}d_{rst}{\bar{\Omega}}_{ij}-{\omega}^{ti}d_{irt}{\bar{\Omega}}^r_j)
   - ({\omega}^{ti}d_{jrt}{\bar{\Omega}}^r_i - {\omega}^{ti}{\omega}_{ij}d_{rst}{\bar{\Omega}}^{rs})=0-0=0  \]
   
In the next Section, we shall revisit these computations in a quite different framework and explain the resulting confusion done between the {\it div} operator induced by the {\it Bianchi} operator and the {\it Cauchy} operator which is the formal adjoint of the {\it Killing} operator [22,23].   \\

\newpage

\noindent
{\bf 2) DIFFERENTIAL DUALITY}  \\

First of all, we describe the initial part of the differential sequence introduced in the preceding Section, calling the successive operators by using their historical names {\it Killing}, {\it Riemann}, {\it Ricci}, {\it Bianchi}, {\it Beltrami}. In particular, lowering the indices by means of the constant metric $\omega$, we obtain:  \\
\[  {\Omega}_{ij}=d_i{\xi}_j+d_j{\xi}_i \Rightarrow   tr(\Omega)=2 d_r{\xi}^r \Rightarrow  R^k_{l,ij}=0 \Leftrightarrow R_{ij}=0 \oplus {\Sigma}^k_{l,ij}=0  \]
and we have exhibited the last splitting allowing to get a direct sum. As a byproduct, we have thus $E_{ij}=0$ and it is well known that the so-called {\it divergence} condition ${\omega}^{ti}d_tE_{ij}=d_tE^t_j=0$ is implied by the Bianchi identities  ${\sum}_{(ijr)} d_rR^k_{l,ij}=0$ where the sum is over the cyclic permutation.  \\
Now we recall that the above differential sequence where {\it Riemann} generates the CC of {\it Killing}, {\it Bianchi} generates the CC of {\it Riemann} and so on, is locally isomorphic to the tensor product of the {\it Poincar\'{e} sequence} by a Lie algebra with $n(n+1)/2$ infinitesimal generators ([11], p 186,224)([21], Section 5). {\it It has therefore a very special property for the formal adjoint operators}, namely that $ad(Riemann)=Beltrami$ generates the CC of $ad(Bianchi)$ while $ad(Killing)=Cauchy$ generates the CC of $ad(Riemann)$, a quite difficult result of {\it homological algebra} saying that the {\it extension modules} of a (differential) module $M$ do not depend on the resolution of $M$ [1,4,6,7,12,13,14,26,27]. Of course, the same property is also valid for the corresponding conformal sequence with now $(n+1)(n+2)/2$ infinitesimal generators whenever $n\geq 3$. The key contradicting results will be provided by the following Theorem and Corollary [12,13,16,20,22]:  \\

\noindent
{\bf THEOREM 2.1}: Contrary to the Ricci operator, the Einstein operator is self-adjoint and we have the following diagram when $n=4$:  \\
\[   \begin{array}{rcccccccccl}
 &4 & \stackrel{Killing}{\longrightarrow} & 10 & \stackrel{Riemann}{\longrightarrow} & 20 & \stackrel{Bianchi}{\longrightarrow} & 20 & \longrightarrow & 6 & \rightarrow 0 \\
  &   &                                                            & \parallel &       & \downarrow  &  & \downarrow &  &   \\
 &    &                                                            &  10     & \stackrel{Einstein}{\longrightarrow}  & 10  & \stackrel{div}{\longrightarrow} & 4    & \rightarrow & 0                      &   \\  
  &   &  &  &  &  &  &  &  \\   
0 \leftarrow & 4 & \stackrel{Cauchy}{\longleftarrow} & 10 & \stackrel{Beltrami}{\longleftarrow} & 20 & \longleftarrow & 20 &  & & \\
                     &           &                                                         & \parallel &    & \uparrow  &  &  &    \\
  &    &      & 10 &  \stackrel{Einstein}{\longleftarrow} & 10 &  &  &  & & 
\end{array}   \]

\noindent
{\it Proof}: The $6$ terms ($4$ for $R_{ij}$ and $2$ for $tr(R)$) are exchanged between themselves by $ad$.  \\
\hspace*{12cm}  Q.E.D.   \\

\noindent
{\bf COROLLARY 2.2}: The Einstein equations in vacuum cannot be parametrized and it is thus not possible to express any generic solution by means of the derivatives of a certain number of arbitrary functions or {\it potentials} like Maxwell equations. \\
\[  \begin{array}{rcccl}
  &  &  &\stackrel{Riemann}{ }  & 20   \\
  & &  & \nearrow &    \\
 4 &  \stackrel{Killing}{\longrightarrow} & 10 & \stackrel{Einstein}{\longrightarrow} & 10  \\
  & & & &  \\
 4 & \stackrel{Cauchy}{\longleftarrow} & 10 & \stackrel{Einstein}{\longleftarrow} & 10 
\end{array}  \]

\noindent
{\it Proof}: According to crucial results of Algebraic Analysis, the test for knowing if a given operator ${\cal{D}}_1$ can be parametrized by an operator ${\cal{D}}$, that is if we can find a differential sequence:
\[   \xi \stackrel{{\cal{D}}}{\longrightarrow} \eta \stackrel{{\cal{D}}_1}{\longrightarrow} \zeta     \]
where ${\cal{D}}_1$ generates the CC of an operator ${\cal{D}}$, has $5$ steps if one uses the identity $ad(ad({\cal{D}}))={\cal{D}}$:  \\
\[   {\cal{D}}_1 \Rightarrow ad({\cal{D}}_1) \Rightarrow ad({\cal{D}}) \Rightarrow ad(ad({\cal{D}}))={\cal{D}} \Rightarrow  {{\cal{D}}_1}' \]
where $ad({\cal{D}})$ generates the CC of $ad({\cal{D}}_1)$ and ${{\cal{D}}_1}'$ generates the CC of ${\cal{D}}$, a parametrization being achieved if and only if ${{\cal{D}}_1}'={\cal{D}}_1$. We obtain therefore the adjoint differential sequence between convenient test functions used in order to construct the various adjoint operators:  \\
\[          \nu  \stackrel{ad({\cal{D}})}{\longleftarrow} \mu  \stackrel{ad({\cal{D}}_1)}{\longleftarrow} \lambda      \]
\hspace*{12cm}   Q.E.D.   \\

We are now ready to explain the results presented in the Introduction. Indeed, with arbitrary test functions $\lambda$, we have:   \\
\[  {\lambda}^{ij}E_{ij}={\lambda}^{ij}(R_{ij}-\frac{1}{2}{\omega}_{ij}tr(R))=({\lambda}^{ij}-\frac{1}{2}{\omega}^{ij} tr(\lambda))R_{ij}=
{\bar{\lambda}}^{ij}R_{ij}   \]
Accordingly, as we just saw that $ad(Eintein)=Einstein$ is parametrizing $ad(Killing)=Cauchy$, then $ad(Ricci)$ is thus also parametrizing 
$Cauchy$ and we obtain through an integration by parts ({\it care} to the following dumb summations):  \\
\[  \begin{array}{rcl}
 2 {\bar{\lambda}}^{ij}R_{ij}  & =  & {\bar{\lambda}}^{ij}{\omega}^{rs}(d_{ij}{\Omega}_{rs}+d_{rs}{\Omega}_{ij}-d_{ri}{\Omega}_{sj} -d_{sj}{\Omega}_{ri})  \\
   &  \equiv & (\Box {\bar{\lambda}}^{rs}+ {\omega}^{rs}d_{ij}{\bar{\lambda}}^{ij}-{\omega}^{sj}d_{ij}{\bar{\lambda}}^{ri}- {\omega}^{ri}d_{ij}{\bar{\lambda}}^{sj}){\Omega}_{rs} \,\,\,\, mod(div)   \\
    &  =  &  {\sigma}^{rs}{\Omega}_{rs}
    \end{array}   \]
{\it Surprisingly}, all the terms after the Dalembertian have already been obtained in the preceding Section and factorize through the divergence operator $d_i{\bar{\lambda}}^{ri}$. Therefore, suppressing the bar for simplicity, we may add the {\it differential constraints} $d_i{\lambda}^{ri}=0$ in a coherent way with the identities:   \\
\[   d_r{\sigma}^{rs}={\omega}^{ij}d_{rij} {\lambda}^{rs}+{\omega}^{rs}d_{rij}{\lambda}^{ij}-
{\omega}^{sj}d_{rij}{\lambda}^{ri} - {\omega}^{ri} d_{rij}{\lambda}^{sj}=0  \]
However, it must be noticed that the potential test functions are arbitrary by definition and can be restricted by such differential constraints as will be shown in the last example of this paper. With more details, we have the identity:  \\
\[   Ricci \circ Killing \equiv 0 \Leftrightarrow ad(Killing)\circ ad(Ricci)\equiv 0   \]
Now, we recall that if ${\cal{D}}$ has coefficients in a differential field $K$ and defines a differential module over the ring $D=K[d]$ of differential operators, we may define the {\it differential transcendence degree} $diff \,\, trd({\cal{D}})=m-rk_D(M)$. We obtain thus [11,12,22]:  \\
\[    diff \,\, trd (Killing)=0 \,\,\Rightarrow  \,\,diff \,\, trd (ad(Ricci))=diff \,\, trd (Ricci)=n(n+1)/2 - n=n(n-1)/2  \]
Taking into account the preceding constraints, we obtain a minimum {\it relative parametrization} that cannot be reduced (See [19,20,21,24,25] for more details and the use of Computer Algebra). \\
Finally, it is important to notice that the {\it div} operator induced by the {\it Bianchi} operator in the upper part of the preceding diagram generates the CC of the {\it Einstein} operator. It follows that the {\it Cauchy} operator does generate the CC of the {\it ad(Einstein)=Einstein} operator in the lower part of the same diagram, {\it though there is no relation at all between these two operators}. It is therefore possible to avoid totally the {\it Einstein} operator which has no mathematical meaning as {\it no specific diagram chasing can produce it} and to keep only the {\it Ricci} operator which has indeed a mathematical meaning only depending on the second order jets ({\it elations}) of the conformal group described by the symbol ${\hat{g}}_2$ through a delicate diagram chasing as we saw previously [9,10,11,16,22]. \\

\noindent
{\bf EXAMPLE 2.3}: We finally provide an elementary but non-trivial example of the methods used and ask the reader to compare the various situations. If $x=(x^1,x^2)$ are the independent variables and $\eta=({\eta}^1,{\eta}^2)$ are the unknowns, let us consider the first order operator ${\cal{D}}_1$ with coefficients in the differential field $K=\mathbb{Q}(x^1,x^2)$ and the same formal notations as before:  \\
\[   d_1{\eta}^1+d_2{\eta}^2 - x^2 {\eta}^1= \zeta   \]
Multiplying by a test function $\lambda$ and integrating by parts, we get $ad({\cal{D}}_1)$ in the form:  \\
\[  \left\{ \begin{array}{lcl}
-d_1\lambda - x^2 \lambda &  =  {\mu}^1  \\
-d_2\lambda  & = {\mu}^2
\end{array} \right.  \,\,\, \Rightarrow \,\,\,  \lambda=d_1{\mu}^2-d_2{\mu}^1+ x^2{\mu}^2      \] 
The generating CC $ad({\cal{D}})$ are:  \\
\[  \left\{   \begin{array}{lcl}
-d_{11}{\mu}^2+d_{12}{\mu}^1-2x^2d_1{\mu}^2+x^2d_2{\mu}^1-(x^2)^2{\mu}^2-{\mu}^1  &  =  &  {\nu}^1  \\
-d_{12}{\mu}^2+d_{22}{\mu}^1-x^2d_2{\mu}^2-2{\mu}^2  &  =  &  {\nu}^2
\end{array}  \right.   \,\,\,  \Rightarrow \,\,  d_1{\nu}^2-d_2{\nu}^1+x^2{\nu}^2=\theta   \]
Multiplying by the test functions $\xi=({\xi}^1,{\xi}^2)$, then adding and integrating by parts, we get the {\it second order} parametrization:  \\
\[  \left\{   \begin{array}{lcl}
 \,\, d_{12}{\xi}^1+d_{22}{\xi}^2-x^2d_2{\xi}^1-2{\xi}^1 &  =  &  {\eta}^1  \\
-d_{11}{\xi}^1-d_{12}{\xi}^2+2x^2d_1{\xi}^1+x^2d_2{\xi}^2-(x^2)^2{\xi}^1-{\xi}^2  &  =  &  {\eta}^2
\end{array}  \right.   \]
and the two differential sequences:  \\
\[   \begin{array}{rcccccccl}
 0 \rightarrow     &   \phi  &  \stackrel{{\cal{D}}_{-1}}{\longrightarrow} & \xi & \stackrel{{\cal{D}}}{\longrightarrow}& \eta & \stackrel{{\cal{D}}_1}{\longrightarrow}  & \zeta  & \rightarrow 0  \\
 0 \leftarrow  & \theta & \stackrel{ad({\cal{D}}_{-1})}{\longleftarrow}  & \nu & \stackrel{ad({\cal{D}})}{\longleftarrow} & \mu & \stackrel{ad({\cal{D}}_1}{\longleftarrow} &  \lambda  & \leftarrow 0  
 \end{array}    \]
 showing that the differential module over $D=K[d_1,d_2]$ defined by ${\cal{D}}_1$ is projective (Exercise).  \\
 Choosing $({\xi}^1=\xi, {\xi}^2=0)$ or $({\xi}^1=0, {\xi}^2={\xi}')$, we obtain two minimal parametrizations but we can also suppose that we add the differential constraint $d_1{\xi}^1+d_2{\xi}^2=0$ in order to obtain the following {\it first order relative parametrization}, a result not evident at first sight [19]:  \\
 \[  \left\{   \begin{array}{lcl}
  -x^2d_2{\xi}^1- 2{\xi}^1  &  =  &  {\eta}^1  \\
 x^2d_1{\xi}^1-(x^2)^2{\xi}^1-{\xi}^2 &  =  &  {\eta}^2
 \end{array}  \right.    \]
We may also set ${\xi}^1=d_2\phi , {\xi}^2= - d_1\phi$ and obtain the new {\it second order} parametrization:  \\
\[  \left\{  \begin{array}{lcl}
-x^2d_{22}\phi-2d_2\phi & = & {\eta}^1  \\
+x^2d_{12}\phi- (x^2)^2d_2\phi+d_1\phi  &  =  &  {\eta}^2
\end{array}  \right.   \]
We finally notice that the choice $\xi={\cal{D}}_{-1}\phi$, namely ${\xi}^1=d_2\phi, {\xi}^2=-d_1\phi + x^2 \phi$ is not allowed as it only provides the trivial solution $\eta=0$.  \\

\newpage

\noindent
{\bf 3) CONCLUSION}  \\

As we have seen, only {\it homological algebra} allows to prove that a differential sequence ${\cal{D}}, {\cal{D}}_1, {\cal{D}}_2$ starting with a Lie operator determined by the action of a lie group on a manifold of dimension $n$ and where each operator generates the CC of the previous one is such that, in the adjoint sequence $ ad({\cal{D}}_2),ad({\cal{D}}_1), ad({\cal{D}})$, each operator generates the CC of the preceding one. This is in particular the case for the first order {\it Killing} operator ${\cal{D}}$, followed by the second order {\it Riemann} operator ${\cal{D}}_1$ and the first order {\it Bianchi} operator ${\cal{D}}_2$. The corresponding part of the adjoint sequence is therefore successively made by 
$ad({\it Bianchi}), ad({\it Riemann})=Beltrami$ and $ad({\it Killing})=Cauchy $. Accordingly, the classical {\it div} operator, induced by the {\it Bianchi} operator and describing the CC of the {\it Einstein} operator, {\it has nothing to do} with the {\it Cauchy} operator. Such a confusion has been produced by the fact that $ad(Einstein)=Einstein$ is thus parametrizing the {\it Cauchy} operator but it is not evident that the transformation of the {\it Einstein } operator described in the Abstract and in the Introduction, just amounts to parametrize the {\it Cauchy} operator / {\it stress} equations by means of the operator {\it ad(Ricci)}. This result is showing that the {\it Einstein} operator is no longer needed and must therefore be taken into account in any future work on gravitational waves.  \\

\vspace{2cm}

\noindent
{\bf REFERENCES}  \\

\noindent
[1] Assem, I.: Alg\`ebres et Modules, Masson, Paris (1997).  \\
\noindent
[2] Eisenhart, L.P.: Riemannian Geometry, Princeton University Press, Princeton (1926).  \\
\noindent
[3] Foster, J., Nightingale, J.D.: A Short Course in General relativity, Longman (1979).  \\
\noindent
[4] Hu,S.-T.: Introduction to Homological Algebra, Holden-Day (1968).  \\
\noindent
[5] Hughston, L.P., Tod, K.P.: An Introduction to General Relativity, London Math. Soc. Students Texts 5, Cambridge University Press 
(1990). \\
\noindent
[6] Kashiwara, M.: Algebraic Study of Systems of Partial Differential Equations, M\'{e}moires de la Soci\'{e}t\'{e} 
Math\'{e}matique de France, 63 (1995) (Transl. from Japanese of his 1970 MasterÕs Thesis).  \\
\noindent
[7] Northcott, D.G.: An Introduction to Homological Algebra, Cambridge university Press (1966).  \\
\noindent
[8] Pommaret, J.-F.: Systems of Partial Differential Equations and Lie Pseudogroups, Gordon and Breach, New York (1978); Russian translation: MIR, Moscow (1983).\\
\noindent
[9] Pommaret, J.-F.: Differential Galois Theory, Gordon and Breach, New York (1983).\\
\noindent
[10] Pommaret, J.-F.: Lie Pseudogroups and Mechanics, Gordon and Breach, New York (1988).\\
\noindent
[11] Pommaret, J.-F.: Partial Differential Equations and Group Theory, Kluwer (1994).\\
http://dx.doi.org/10.1007/978-94-017-2539-2    \\
\noindent
[12] Pommaret, J.-F.: Partial Differential Control Theory, Kluwer, Dordrecht (2001).\\
\noindent
[13] Pommaret, J.-F.: Algebraic Analysis of Control Systems Defined by Partial Differential Equations, in "Advanced Topics in Control Systems Theory", Springer, Lecture Notes in Control and Information Sciences 311 (2005) Chapter 5, 155-223.\\
\noindent
[14] Pommaret, J.-F.: Parametrization of Cosserat Equations, Acta Mechanica, 215 (2010) 43-55.\\
http://dx.doi.org/10.1007/s00707-010-0292-y  \\
\noindent
[15] Pommaret, J.-F.: Spencer Operator and Applications: From Continuum Mechanics to Mathematical Physics, in "Continuum Mechanics-Progress in Fundamentals and Engineering Applications", Dr. Yong Gan (Ed.), ISBN: 978-953-51-0447--6, InTech, 2012, Available from: \\
http://dx.doi.org/10.5772/35607   \\
\noindent
[16] Pommaret, J.-F.: The Mathematical Foundations of General Relativity Revisited, Journal of Modern Physics, 4 (2013) 223-239. \\
 http://dx.doi.org/10.4236/jmp.2013.48A022   \\
 \noindent
[17] Pommaret, J.-F.: The Mathematical Foundations of Gauge Theory Revisited, Journal of Modern Physics, 5 (2014) 157-170.  \\
http://dx.doi.org/10.4236/jmp.2014.55026  \\
\noindent
[18] Pommaret, J.-F.: From Thermodynamics to Gauge Theory: the Virial Theorem Revisited, pp. 1-46 in "Gauge Theories and Differential geometry,", NOVA Science Publisher (2015).  \\
\noindent
[19] Pommaret, J.-F.: Relative parametrization of Linear Multidimensional Systems, Multidim. Syst. Sign. Process., 26 (2015) 405-437.  \\
DOI 10.1007/s11045-013-0265-0   \\
\noindent
[20] Pommaret, J.-F.: Airy, Beltrami, Maxwell, Einstein and Lanczos Potentials revisited, Journal of Modern Physics, 7, 699-728 (2016). \\
\noindent
http://dx.doi.org/10.4236/jmp.2016.77068   \\
\noindent
[21] Pommaret, J.-F.: Deformation Theory of Algebraic and Geometric Structures, Lambert Academic Publisher (LAP), Saarbrucken, Germany (2016). \\
http://arxiv.org/abs/1207.1964  \\
\noindent
[22] Pommaret, J.-F.: Algebraic Analysis and Mathematical Physics (2017).  \\
https://arxiv.org/abs/1706.04105   \\
\noindent
[23] Pommaret, J.-F.: Differential Algebra and Mathematical Physics (2017).  \\
https://arxiv.org/abs/1707.09763   \\
\noindent
[24] Pommaret, J.-F. and Quadrat, A.: Localization and Parametrization of Linear Multidimensional Control Systems, Systems \& Control Letters, 37 (1999) 247-260.  \\
\noindent
[25] Quadrat, A.: An Introduction to Constructive Algebraic Analysis and its Applications, 
Les cours du CIRM, Journees Nationales de Calcul Formel, 1(2), 281-471 (2010).\\
\noindent
[26] Rotman, J.J.: An Introduction to Homological Algebra, Pure and Applied Mathematics, Academic Press (1979).  \\
\noindent
[27] Schneiders, J.-P.: An Introduction to D-Modules, Bull. Soc. Roy. Sci. Li\`{e}ge, 63 (1994) 223-295.  \\
\noindent
[28] Spencer, D.C.: Overdetermined Systems of Partial Differential Equations, Bull. Am. Math. Soc., 75 (1965) 1-114.\\
\noindent
[29] Vessiot, E.: Sur la Th\'{e}orie des Groupes Infinis, Ann. Ec. Norm. Sup., 20 (1903) 411-451.  \\
(Can be obtained from http://numdam.org).  \\

\end{document}